\documentclass[twocolumn,rmp,nofootinbib,floatfix,superscriptaddress,showkeys]{revtex4}

\usepackage{dcolumn}
\usepackage{graphicx}
\usepackage{rotating}
\usepackage{amsmath,amsfonts,amssymb}

\newcommand{\s}{f}
\newcommand{\kg}{\,\mathrm{kg}}
\newcommand{\g}{\,\mathrm{g}}
\newcommand{\My}{\,\mathrm{My}}
\newcommand{\e}{\mathrm{e}}

\newcommand{\xmin}{\ensuremath{x_{\min}}}
\newcommand{\mmin}{\ensuremath{M_{\min}}}

\newcommand{\remove}[1]{}

\begin{document}

\title{How many species have mass $M$?\footnote{This manuscript is a pre-print version that has not undergone final editing.  Please refer to the complete version of record in the {\em American Naturalist} (2008) at \\ {\tt http://www.journals.uchicago.edu/toc/an/current}.  }}
\author{Aaron Clauset}
\email{aaronc@santafe.edu}
\affiliation{Santa Fe Institute, 1399 Hyde Park Road, Santa Fe NM, 87501, USA}
\author{David J. Schwab}
\email{djs47@physics.ucla.edu}
\affiliation{Department of Physics and Astronomy, University of California, Los Angeles, Los Angeles, CA, 90024, USA}
\author{Sidney Redner}
\email{redner@bu.edu}
\affiliation{Center for Polymer Studies and Department of Physics, Boston University, Boston, MA 02215, USA}
\affiliation{Santa Fe Institute, 1399 Hyde Park Road, Santa Fe NM, 87501, USA}


\begin{abstract}
  Within large taxonomic assemblages, the number of species with adult body 
  mass $M$ is characterized by a broad but asymmetric distribution, with the
  largest mass being orders of magnitude larger than the typical mass. This
  canonical shape can be explained by cladogenetic diffusion that is bounded
  below by a hard limit on viable species mass and above by extinction risks that
  increase weakly with mass. Here we introduce and analytically solve a
  simplified cladogenetic diffusion model. When appropriately parameterized,
  the diffusion-reaction equation predicts mass distributions that are in
  good agreement with data on 4002 terrestrial mammal from the late 
  Quaternary and 8617 extant bird species. Under this model, we show that a 
  specific tradeoff between the strength of within-lineage drift toward larger
  masses (Cope's rule) and the increased risk of extinction from increased
  mass is necessary to produce realistic mass distributions for both taxa. We then 
  make several predictions about the evolution of avian species masses.
\end{abstract}


\maketitle

\section*{Introduction}

For a large taxonomic group under stable macroevolutionary conditions, how many species have a mass $M$? This question has wide implications for the evolution and distribution of many other species characteristics that correlate strongly with body mass, including life span, life history, habitat, metabolism and extinction risk~\citep{calder:1984,brown:1995,bennett:owens:1997,cardillo:etal:2005}.

Extant species, including mammals, birds, fish and insects, exhibit a canonical form of the species mass distribution~\citep{stanley:1973,kozlowski:gawelczyk:2002,allen:etal:2006}, in which the typical mass is an intermediate value; for example, in mammals the typical mass is that of the common Pacific Rat ({\em Rattus exulans}, $40\g$). Larger or smaller species, in turn, are significantly less common, but asymmetrically so: the largest species, such as the extinct Imperial Mammoth ({\em Mammuthus imperator}, $10^{7}\g$) for terrestrial mammals, are many orders of magnitude larger, while the smallest is only a little smaller, e.g., the Remy's Pygmy Shrew ({\em Suncus remyi}, $1.8\g$).

The ubiquity of this distribution of species masses (Fig.~\ref{fig:schematic}a) suggests the existence of a universal evolutionary mechanism.  A theoretical explanation  of this distribution may shed light both on the interaction between ecological and macroevolutionary processes~\citep{stanley:1975}, and on long-term trends in species mass~\citep{alroy:2000a,alroy:2000b}, including Cope's rule, the oft-studied notion that species mass tends to increase within a lineage over evolutionary time~\citep{stanley:1973,alroy:1998}.  Clauset and Erwin~(\citeyear{clauset:erwin:2008}) recently showed, by comparing extensive computer simulations with empirical data, that cladogenetic diffusion in the presence of a taxon-specific lower limit on mass $\mmin$ and extinction risks that grow weakly with mass, can explain both the canonical form described above, and the precise form of the distribution of terrestrial mammal masses.

Here, we present a simplified three-parameter version of the Clauset and Erwin (CE) model for the species mass distribution and solve this model in the steady state. Comparing the predictions of this simplified model with species mass data for the same 4002 terrestrial mammals from the late Quaternary~\citep{smith:etal:2003} (henceforth denoted ``Recent'' mammals, which includes species that have become extinct during the Holocene), we reproduce Clauset and Erwin's results.  We then show that the model's predictions, when appropriately parameterized, are also in good agreement with data for 8617 extant avian species~\citep{dunning:2007}.

\section*{The Clauset-Erwin Model}

\begin{figure}[t]
\begin{center}
\includegraphics[scale=0.43]{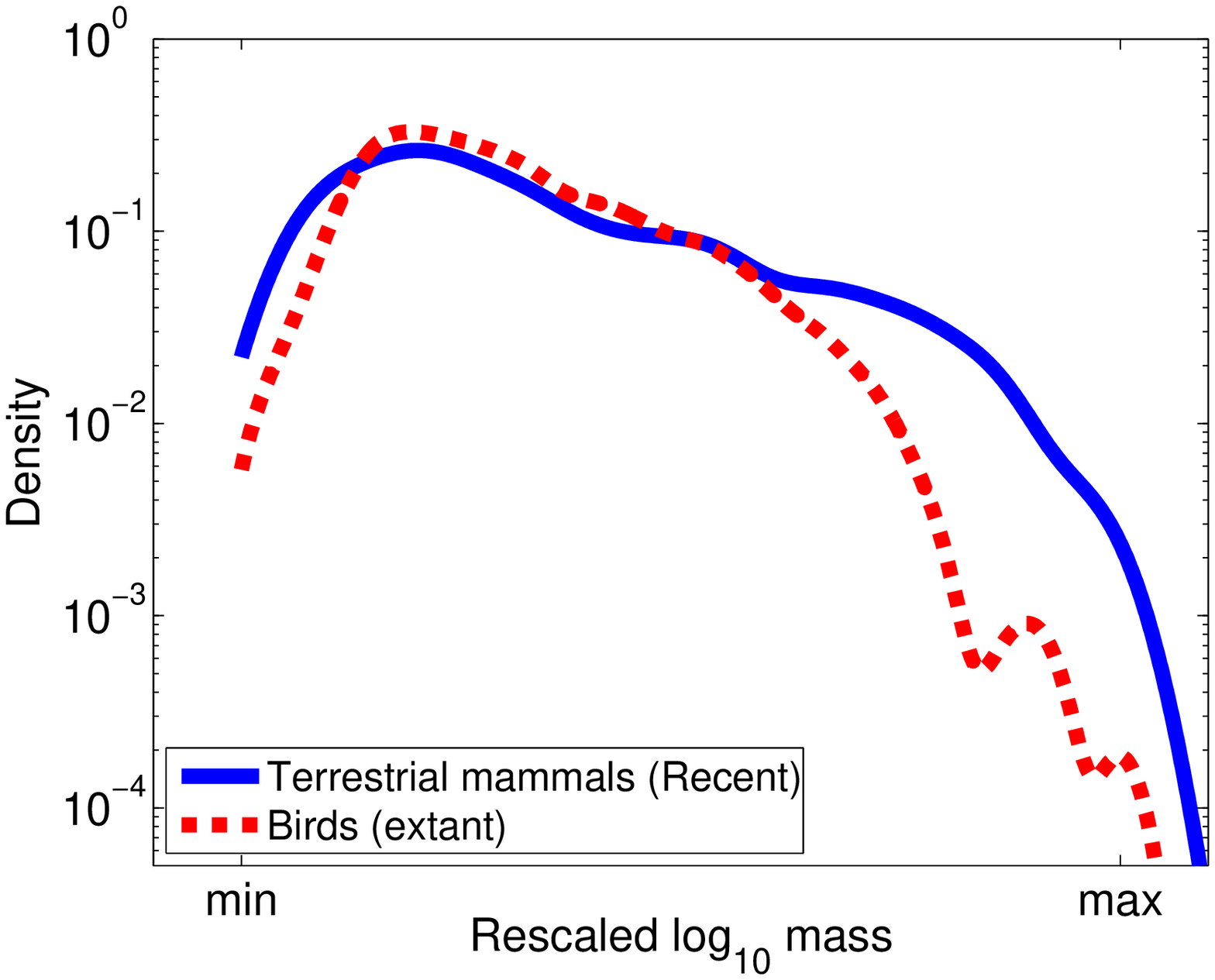} 
\includegraphics[scale=0.69]{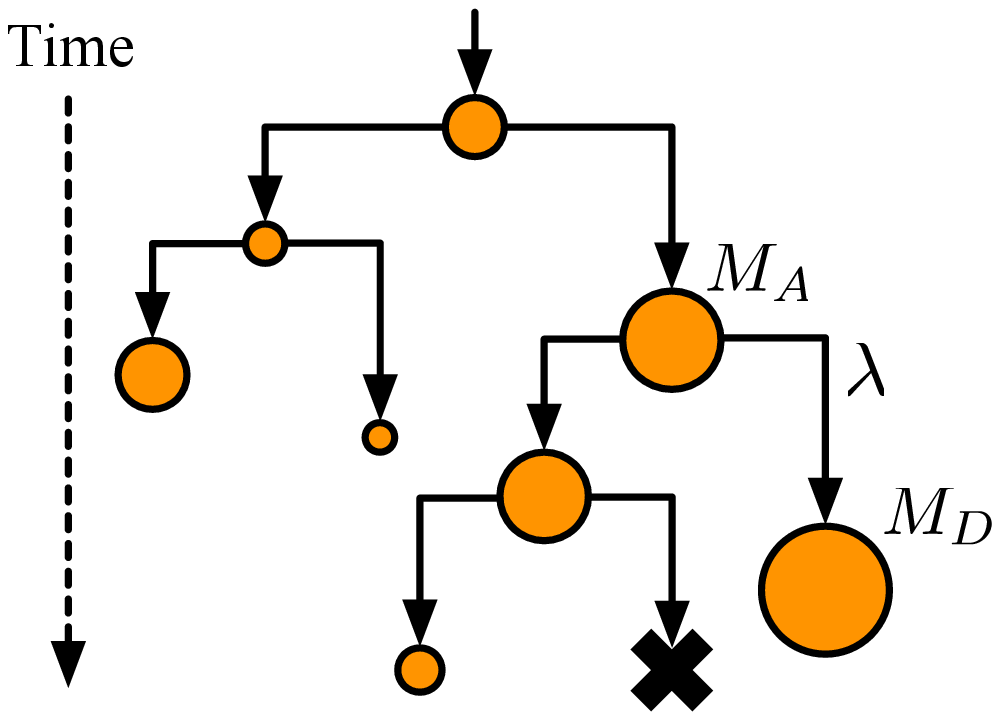} 
\end{center}
\caption{
({\bf a}) 
An illustration of the similarity of the body mass distributions for different taxonomic groups, using data for 4002 Recent terrestrial mammals~\citep{smith:etal:2003} and 8617 extant birds~\citep{dunning:2007}. For visual clarity, the empirical distributions have been smoothed using a Gaussian kernel~\citep{wasserman:2006} and rescaled to lay on the same abscissa interval.
({\bf b}) 
A schematic showing the basic cladogenetic diffusion process by which species mass varies over evolutionary time. The masses of a descendant $M_{D}$ is related to the mass of its ancestor $M_{A}$ by a random multiplicative factor $\lambda$, which represents the total selective influence on species mass from all sources, i.e., $M_{D} = \lambda\,M_{A}$.
}
\label{fig:schematic}
\end{figure}

Many models of the variation of species body mass over evolutionary time assume a cladogenetic diffusion process~\citep{stanley:1973,mckinney:1990,mcshea:1994,kozlowski:gawelczyk:2002} where each descendant species' mass $M_{D}$ is related to its ancestor's mass $M_{A}$ by a random multiplicative factor $\lambda$, i.e., $M_{D}=\lambda\,M_{A}$ (Fig.~\ref{fig:schematic}b), where $\lambda$ represents the total selective influence on the descendant species' mass from all sources. Clauset and Erwin~(\citeyear{clauset:erwin:2008}) studied a family of such diffusion models, whose form and parameters could be estimated empirically. Under their model, the evolution of species mass is bounded on the lower end by hard physiological limits, perhaps from metabolic~\citep{pearson:1948,pearson:1950,west:etal:2002} or morphological constraints~\citep{stanley:1973}, and on the upper end by an extinction risk that increases weakly with mass~\citep{liow:etal:2008}. Near the lower limit, however, within-lineage changes to body mass become increasingly biased toward larger masses, i.e., $\langle\ln\lambda\rangle$ increases as $M\to\mmin$.

Using model parameters estimated from ancestor-descendant data on extinct North American terrestrial mammals since the Cretaceous-Tertiary boundary~\citep{alroy:2008}, Clauset and Erwin simulated $60\My$ of mammalian body mass evolution and found that the predicted distribution closely matches that of 4002 Recent terrestrial mammal species. They also found that simpler diffusion models, e.g., those that omitted either the lower limit, an increased extinction risk from increased mass, or the increased bias toward larger masses as $M\to\mmin$, predicted significantly different distributions. Thus, these three mechanisms can explain not only the canonical form of the species mass distribution, but also its taxon-specific shape.

\section*{A Mathematically Solvable Version}

The CE model, however, remains too complex for mathematical analysis, even though it omits many ecological and microevolutionary processes such as interspecific competition, predation, and population dynamics. By simplifying its assumptions slightly, we can formulate a diffusion-reaction model of species body mass evolution, which retains most of the key features of the CE model and which can be mathematically analyzed. This model can then be used to make inferences about body mass evolution without resorting to laborious simulations.

Let $\s(x,t)$ denote the number of species with logarithmic mass $x=\ln M$ at time $t$, i.e., we transform the CE model's multiplicative diffusion process into an additive diffusion process on a logarithmic scale. The cladogenesis mechanism leads to two offspring species in which \mbox{$x\to x+\ln\lambda$}. In the continuum limit, a value $\langle\ln\lambda\rangle>0$ (Cope's rule) corresponds to positive drift velocity in the equation of motion for $\s(x,t)$.  In the CE model for terrestrial mammals, the drift velocity was estimated from fossil data to increase like $M^{1/4}$ near the lower limit on species mass, a feature that Clauset and Erwin found was necessary to accurately predict the number of small-bodied mammals. (Whether this quarter-power form is related to the quarter-power scaling commonly found elsewhere in the body-size literature~\citep{savage:etal:2004} remains to be seen.) For mathematical simplicity, however, we will ignore this dependence at the expense of possibly mis-estimating the number of small-bodied species. We also assume that selection pressures on body mass are roughly independent, implying that the distribution of changes $\lambda$ is approximately lognormal (but see~\citep{clauset:erwin:2008}).

The feature that two offspring are produced at each update step corresponds to a population growth term in the equation of motion that is proportional to $\s$ itself.  The extinction probability may also be represented by a loss term that is proportional to $\s$.  For terrestrial mammals, recent empirical studies~\citep{liow:etal:2008} support the assumption that the probability per unit time of a species becoming extinct $p_{e}(x)$ grows weakly with its mass. The CE model uses a simple parameterization of this behavior: \mbox{$p_{e}(x)=A+Bx$} where $B\geq0$, which corresponds to an extinction probability that grows logarithmically with species mass. 

Combining these three elements --- diffusion, cladogenesis, and extinction --- we may write  the continuum equation of motion $\s(x,t)$ for the number of species with mass $x=\ln M$ at time $t$ as
\begin{equation}
\label{eq:cxt}
\frac{\partial \s}{\partial t} + v \frac{\partial \s}{\partial x} = D \frac{\partial^2 \s}{\partial x^2} + (1-A-Bx)\s \enspace ,
\end{equation}
where $v=\langle\ln\lambda\rangle$ is the drift velocity (strength of Cope's rule) and the variance $D=\langle\,\ln^{2}\lambda\,\rangle$ is the diffusion coefficient.  (In physics, Eq.~\eqref{eq:cxt} is called the convection-diffusion equation or the Fokker-Planck equation.)  This equation, however, omits the lower limit on species body mass, which we incorporate momentarily.

The time-dependent equation of motion itself may be useful for studying evolutionary trends in species body mass, or for making inferences about correlated extinction or speciation events. For our purposes, however, we are mainly interested in its stationary solution. Such a steady-state should exist whenever all species within the taxon experience roughly the same set of macroevolutionary selective pressures, i.e., under stable macroevolutionary conditions. To derive this solution, we set the time derivative in Eq.~\eqref{eq:cxt} to zero to obtain $\s'' - \mu \s' +(\alpha-\beta x) \s =0$, where $\mu=v/D$, $\alpha=(1-A)/D$, $\beta=B/D$, and the prime denotes differentiation with respect to $x$.  We now eliminate the first derivative term by introducing $\s = \e^{\mu x/2}\,\psi$ to transform the steady-state equation to
\begin{equation*}
  \psi'' +\left[\left(\alpha-\frac{\mu^2}{4}\right)-\beta x\right]\psi =0 \enspace .
\end{equation*}
This equation can be brought into the form of the standard Airy's differential equation~\citep{abramowitz:stegun:1972}
\begin{equation}
\label{eq:ai}
\psi'' -z\psi=0 \enspace ,
\end{equation}
where we introduce the new variable $z=\beta^{1/3}x-\beta^{-2/3}\left(\alpha-{\mu^2}/{4}\right)$ and the prime now denotes differentiation with respect to $z$.  The general solution to Eq.~\eqref{eq:ai} is $\psi(z) = c_1\, {\rm Ai}(z) +c_2\, {\rm Bi}(z)$, where Ai$(z)$ and Bi$(z)$ are the Airy functions.  Since there can be no species with infinite mass we may set $c_2=0$. The parameter $c_1$ is then determined by the normalization of $\s(x)$.

Thus we can now write the species mass distribution as:
\begin{equation}
  \s(x) \propto \e^{{\mu x}/2}\,\, {\rm Ai}\!\left[ \beta^{{1}/{3}} x -
    \beta^{-{2}/{3}}\left(\alpha-\frac{\mu^{2}}{4}\right)\right] \enspace .
\label{eq:result}
\end{equation}

\noindent Including now the taxon-specific lower limit $\xmin$ on species mass implies the constraint $\s(\xmin)=0$ for the steady-state solution, and allows us to eliminate one parameter from Eq.~\eqref{eq:result}.  Using the fact that the first zero of the Airy function is located at \mbox{$z_{0}=-2.3381\!\ldots$}\,, which we now require to coincide with $x=\xmin$, gives the constraint
\begin{displaymath}
z_{0} = \beta^{1/3}\xmin-\beta^{-2/3}\left(\alpha-\mu^{2}/4\right) \enspace, \nonumber
\end{displaymath}
which we may solve for $\alpha$.  Inserting this result into Eq.~\eqref{eq:result} yields
\begin{equation}
\s(x) \propto \e^{\mu x/2}\, {\rm Ai}\left[ \beta^{1/3}(x-x_{min})+z_0\right] \enspace ,
\label{eq:simple}
\end{equation}
as the steady-state solution for the species mass distribution.

If the lower limit $\xmin$ is known, this simple-minded model has only two parameters: $\mu$, associated with the biased diffusion process and $\beta$, associated with the extinction process.  A positive bias in the diffusion $\mu>0$ (Cope's rule) has several systematic effects on the distribution of species masses: it (i) pushes the left tail of the distribution away from the lower limit at $\xmin$, (ii) shifts the modal mass toward slightly larger values, and (iii) extends the right tail of the distribution.  In contrast, increasing $\beta$ implies that species become extinct with greater probability for a given mass $M$, which contracts the right tail of the distribution.  For a given bias $\mu$ and number of species $n$, the parameter $\beta$ also set an effective upper limit on the expected maximum observed mass within the taxon without invoking a hard boundary, e.g., from biomechanical constraints~\citep{mcmahon:1973}. 

\section*{Mammalian Body Mass Evolution}

We now test the predictions of this simple mathematical model using empirical data for 4002 Recent terrestrial mammals~\citep{smith:etal:2003} and 8617 birds~\citep{dunning:2007}. In the former case, we take \mbox{$\mmin=1.8\g$}, the size of the smallest known mammal, and we estimate \mbox{$v=0.109\pm0.021$} and \mbox{$D=0.508\pm0.027$} (SE) from Alroy's ancestor-descendant data for North American terrestrial mammals~\citep{alroy:2008}. Incorporating these values into Eq.~\eqref{eq:simple} leaves only $\beta$ unspecified. A strong test of this model would estimate $\beta$ from fossil data; however, while studies of extinction among mammals suggest that $\beta>0$~\citep{liow:etal:2008}, current data does not appear to be sufficiently detailed to give a precise estimate for mammals. Instead, following Clauset and Erwin~(\citeyear{clauset:erwin:2008}), we choose $\beta$ by minimizing the tail-weighted Kolmogorov-Smirnov (wKS) goodness-of-fit statistic~\citep{press:etal:1992} for the predicted and empirical distributions:
\begin{displaymath}
{\rm wKS} = \max_{x} \frac{\left| S(x) - P(x) \right|}{\sqrt{P(x)(1-P(x)}} \enspace ,
\end{displaymath}
where $S(x)$ is the empirical distribution function and $P(x)$ is the predicted cumulative distribution function. Thus, small values of wKS correspond to a model that is statistically close to the empirical data everywhere. We find that two alternative methods of choosing $\beta$, by numerically matching the modal masses of the model and the empirical data or by matching the expected maximum mass of the model with the observed maximum in the empirical data, produce similar results.

\begin{figure}[t]
\begin{center}
\includegraphics[scale=0.43]{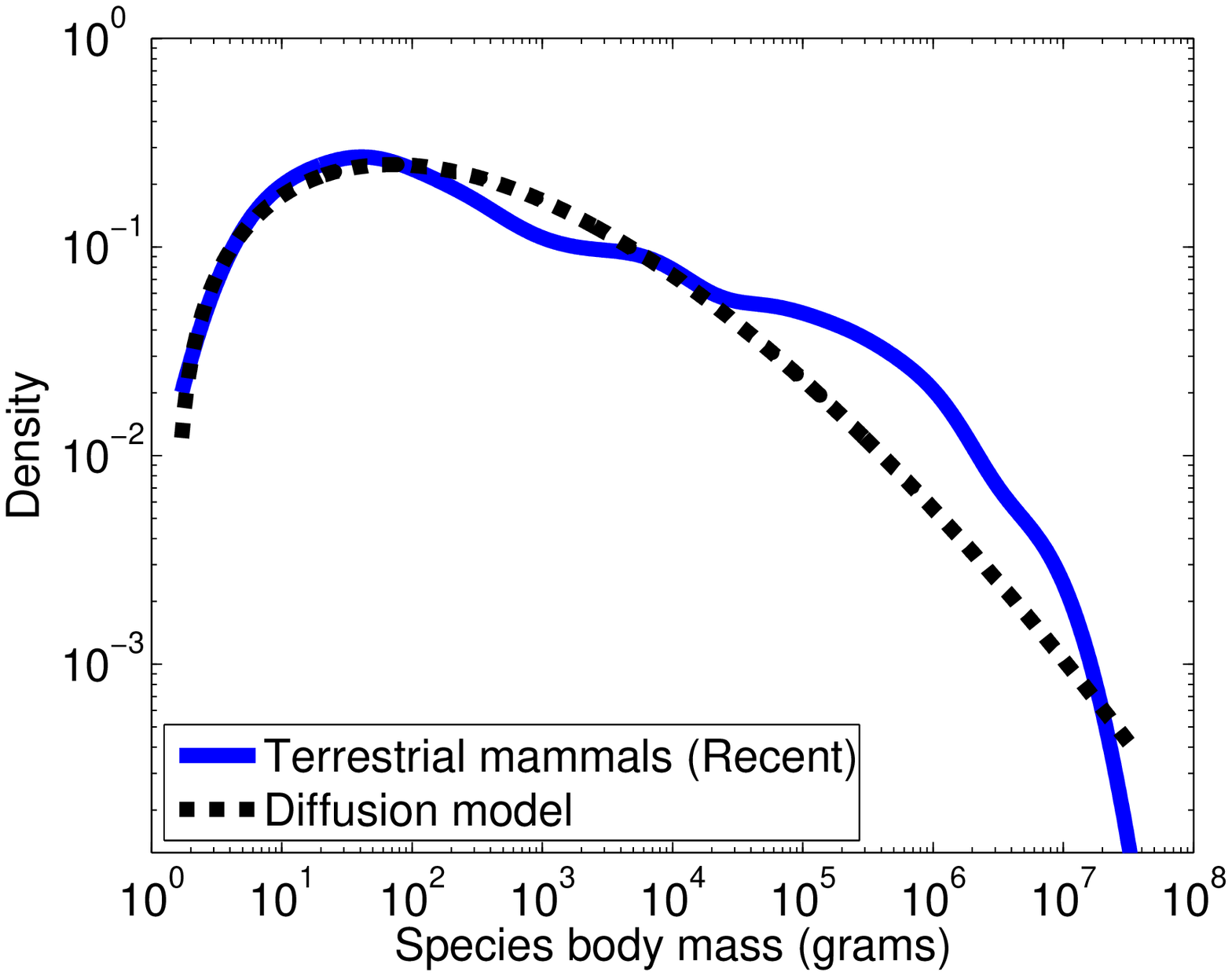} 
\includegraphics[scale=0.43]{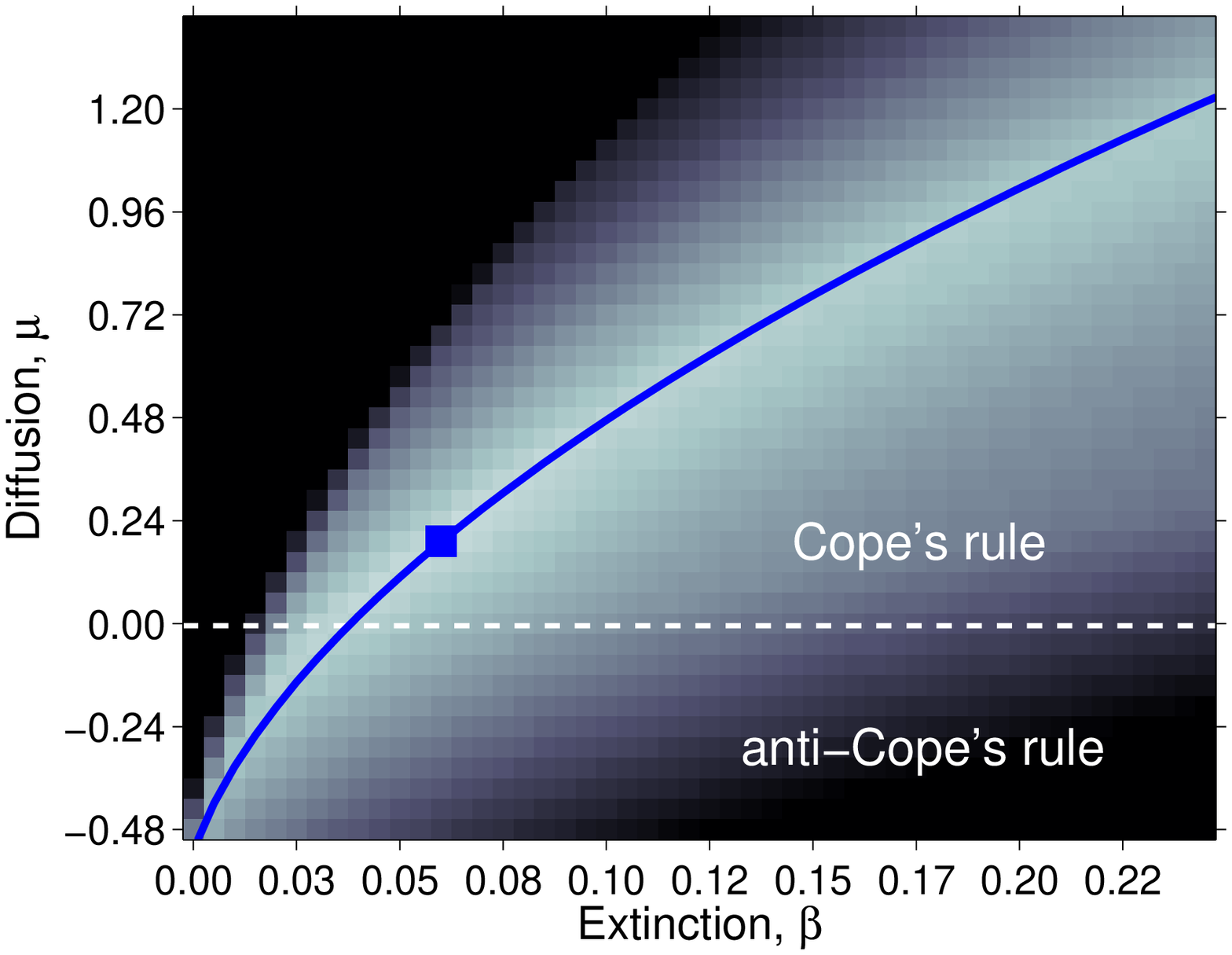} 
\end{center}
\caption{
({\bf a}) 
The solution of the diffusion-reaction model of species body mass evolution [Equation~\eqref{eq:simple}] with (smoothed) empirical data for 4002 terrestrial mammals from the late Quaternary. Parameters $\mu$ and $\xmin$ were estimated from empirical data (see text) on extinct terrestrial mammals since the Cretaceous-Tertiary boundary while $\beta$ was chosen to minimize the tail-weighted Kolmogorov-Smirnov (wKS) distance between the model and the empirical distribution. 
({\bf b}) 
The $(\mu,\beta)$-plane of wKS distance of the specified model and the empirical data for terrestrial mammals (lighter values correspond to smaller wKS). The distinctive groove (blue line) demonstrates that a systematic relationship between the bias parameter $\mu$ and the extinction parameter $\beta$ is necessary to produce realistic mass distributions (small wKS). The parameter pair used in ({\bf a}) is marked. Dashes demarcate the line of no within-lineage bias ($\mu=0$).
}
\label{fig:mammals}
\end{figure}

For the interested reader, details on the preparation of the empirical data are discussed at length by Alroy~(\citeyear{alroy:1998}) and Smith et al.~(\citeyear{smith:etal:2003}) for mammalian fauna, and by Dunning Jr.~(\citeyear{dunning:2007}) for avian fauna. In general, body mass estimates were derived using conventional techniques (for example, dentition techniques for mammals~\citep{damuth:1990}). For simplicity, differences due to sexual dimorphism, geographic variation, etc.\ were ignored or averaged out. Although such differences can be critical for smaller studies, given the scale of our data, in terms of the number of species studied and the wide range of body masses, mild misestimates of body masses are unlikely to change our conclusions unless they are widespread and systematic.

The resulting fit (Fig.~\ref{fig:mammals}a) is in good agreement with the empirical data, except for a slight overestimate of the number of species with mass near $1\kg$, an underestimate of the number near $300\kg$, and a slight misestimate of the number of very small-bodied species. The deviations in the right tail are also seen in the CE model and may be due to, e.g., phylogenetically correlated speciation or extinction events in the recent past. The deviations in the left tail may be due to our omission of the mass-dependence in the drift term $\mu$ identified by Clauset and Erwin; however, incorporating this behavior into our diffusion-reaction model is technically non-trivial.

Thus, the evolution of mammalian species body masses can largely be viewed as a simple diffusion process, characterized by (i) a slight within-lineage drift toward larger masses over evolutionary time (Cope's rule), (ii) a hard lower boundary on how small body masses can become, and (iii) a very soft constraint on large body masses in the form of increased extinction risk. Phrased more conceptually, the left tail of the mammalian body mass distribution is mainly controlled by the lower limit on mass, while the right tail is the result of an evolutionary tradeoff at different timescales: over the short-term, within-lineage increases in body mass offer selective advantages such as better tolerance of resource fluctuations, better thermoregulation, better predator avoidance, etc.~\citep{calder:1984,brown:1995}, while they also increase the long-term risk of extinction --- a tradeoff previously identified in the more specific case of carnivorous mammals~\citep{valkenburgh:1999,valkenburgh:etal:2004}.

\section*{Avian Body Mass Evolution}

\begin{figure}[t]
\begin{center}
\includegraphics[scale=0.43]{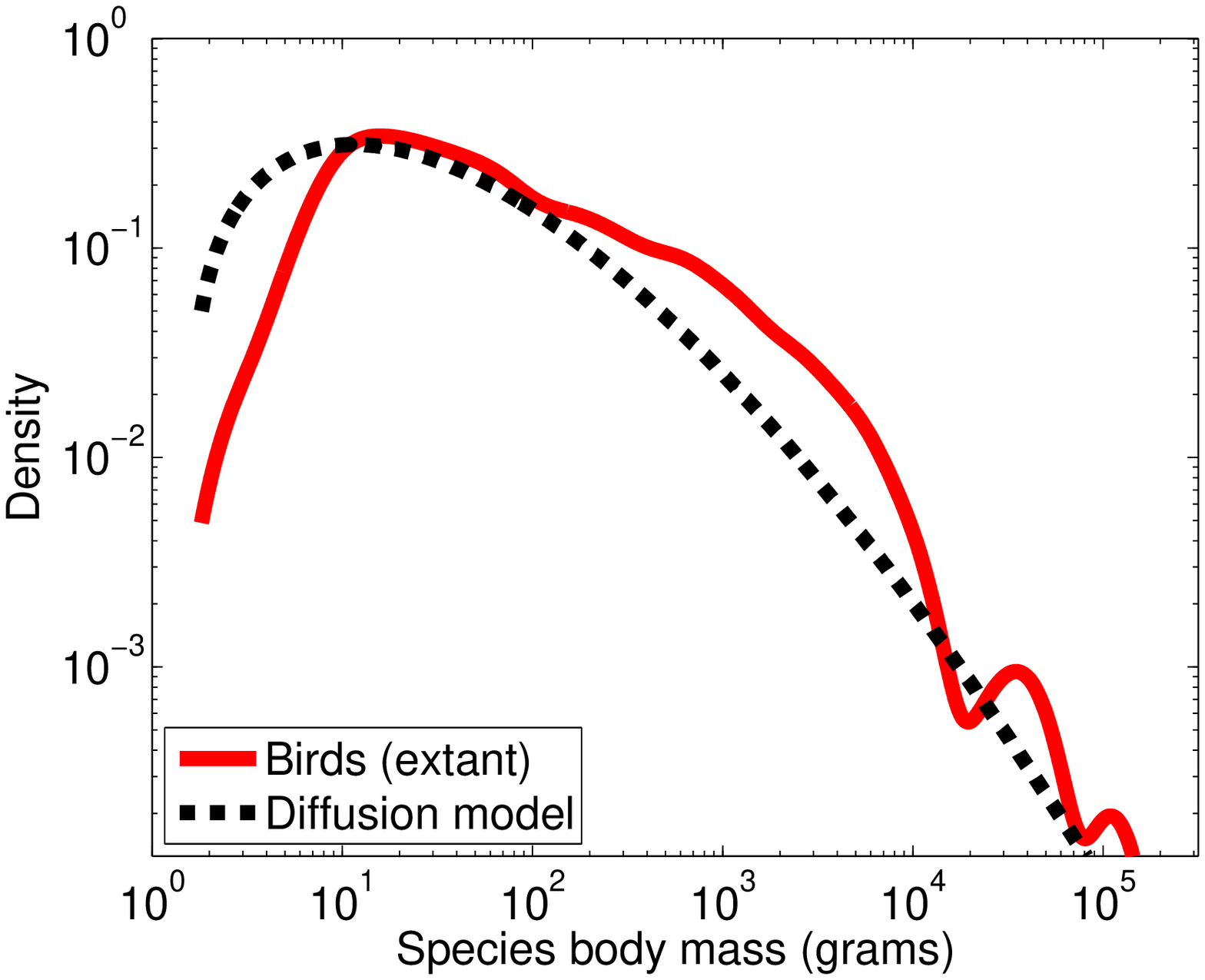} 
\includegraphics[scale=0.43]{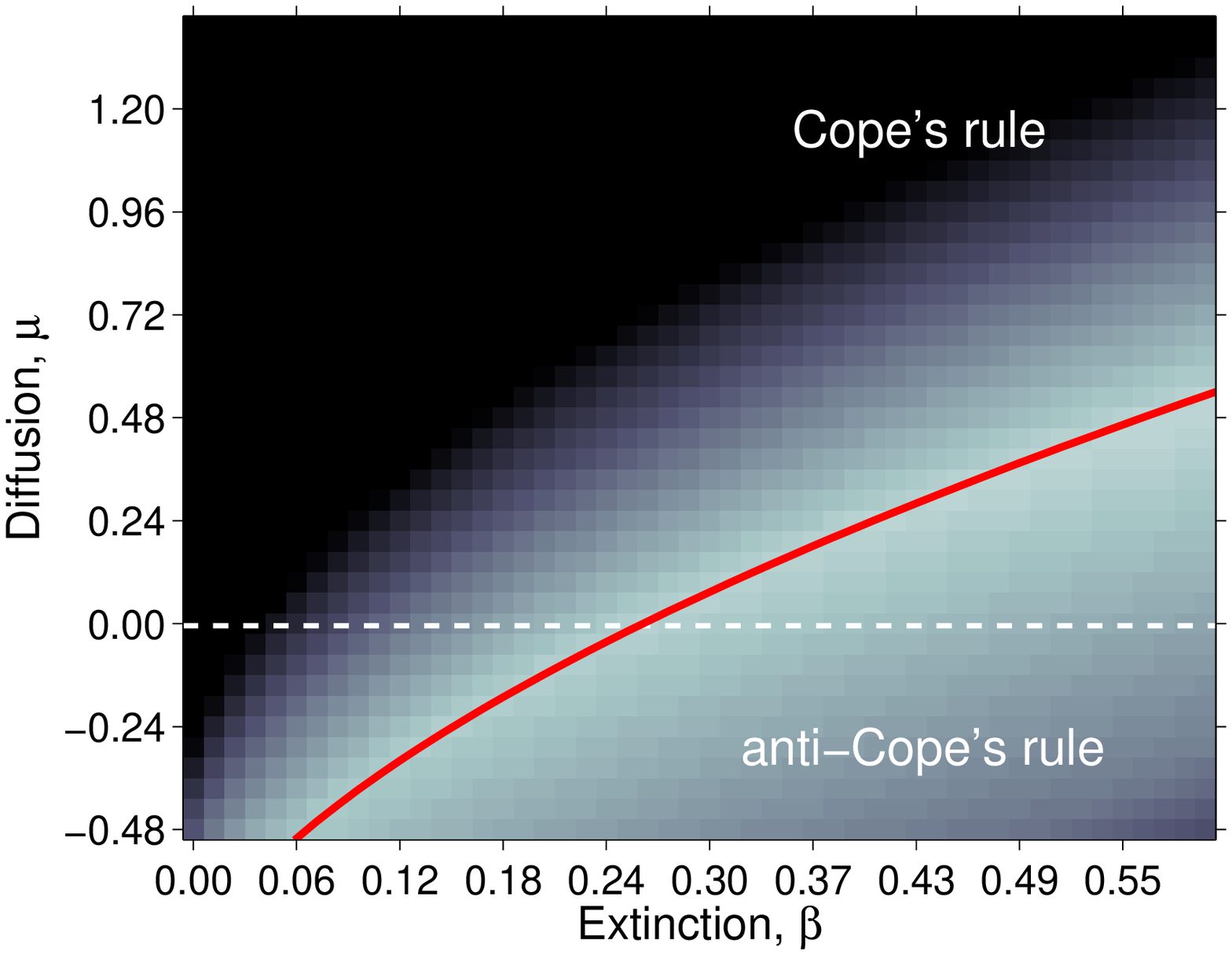}
\end{center}
\caption{
({\bf a}) 
The steady-state solution of the diffusion-reaction model of species body mass evolution [Equation~\eqref{eq:simple}], with $\mu=0$, $\mmin=2\g$ and $\beta$ chosen to minimize the tail-weighted Kolmogorov-Smirnov (wKS) distance between the model and the empirical distribution, with (smoothed) empirical data for 8617 avian species. 
({\bf b}) 
The $(\mu,\beta)$-plane of wKS distances (lighter values correspond to smaller wKS), showing a systematic relationship between $\mu$ and $\beta$ for producing realistic (small wKS) avian mass distributions. The form of this relationship is similar to that of mammals (Fig.~\ref{fig:mammals}b), suggesting a similar overall macroevolutionary mechanism.
}
\label{fig:birds}
\end{figure}

Unlike mammals, data on most other taxonomic groups are generally not sufficient to yield accurate estimates of the parameters $\mu$ and $\beta$ (but see Novack-Gottshall and Lanier~(\citeyear{novack-gottshall:lanier:2008})). The evolutionary history of mammalian body masses is relatively clear, in part because mammalian fossils are relatively plentiful, are often sufficiently well-preserved that body mass estimates can be made~\citep{valkenburgh:1990,damuth:1990}, and the distribution of species body masses during an apparently stable evolutionary period is known. Avian species, however, present an interesting case for study using our model: the distribution of extant avian body masses (Fig.~\ref{fig:birds}a) is relatively well characterized~\citep{dunning:2007} and evidence of a minimum species body mass $\mmin$ is reasonable~\citep{pearson:1950}. However, the avian fossil record may be too sparse to yield accurate estimates of $\mu$ and $\beta$~\citep{fountaine:etal:2005,hone:etal:2008}.

Even without estimates of $\mu$ and $\beta$, however, the diffusion model can be used to make quantitative statements about the general character of avian body mass evolution. To demonstrate this, we consider which combinations of the parameters $\mu,\beta$ produce ``realistic'' mass distributions, i.e., those with a small distributional distance to the empirical distribution. In particular, we compute the wKS distance between the model and the empirical data over the $(\mu,\beta)$-plane and determine the regions that yield the best fits. To illustrate this technique in a better understood context, we first apply it to the data on Recent terrestrial mammals, disregarding for the moment that we have an estimate of $\mu$ from fossil data.

The result of this exercise (Fig.~\ref{fig:mammals}b) shows that realistic mammalian mass distributions can be produced by a wide, but not arbitrary, variety of biases $\mu$ and extinction risks $\beta$, including no bias at all, i.e., $\mu=0$. This degeneracy, which forms a groove in the $(\mu,\beta)$-plane following roughly $\mu\propto\beta^{\theta}$ with $\theta\approx0.5$, suggests that part of the difficulty in determining for a particular taxon whether mass evolution is biased toward larger sizes or not (see for example~\citep{maurer:etal:1992,maurer:1998,bokma:2002}) is that a positive bias is not a necessary condition for the evolution of realistic mass distributions. Indirect tests of the sign of $\mu$, based either on the mass distribution within subclades~\citep{mcshea:1994,wang:2001} or on changes to minimum and maximum masses within a subclade over geologic time~\citep{jablonski:1997} may be adequate if confounding hypotheses can be eliminated or if an appropriate null model is available. On the other hand, the most accurate method remains the direct analysis of large amounts of ancestor-descendant data~\citep{alroy:1998,alroy:2000a,novack-gottshall:lanier:2008}, preferably when derived from as realistic a phylogeny as possible. An alternative approach, however, could use the systematic relationship between $\mu$ and $\beta$ (Fig.~\ref{fig:mammals}b) to infer $\mu$ from an empirical estimate of $\beta$.

Performing the same analysis with the birds data (Fig.~\ref{fig:birds}b), we find that realistic avian mass distributions are also only produced by particular combinations of parameters, and also that the groove of minimum wKS values follows a systematic relationship with the same basic form as that of terrestrial mammals (Fig~\ref{fig:mammals}b), i.e., $\mu\propto\beta^{\theta}$ with $\theta\approx0.5$. Like mammals, this groove passes through the point $\mu=0$, implying that this test cannot rule out the possibility that Cope's rule ($\mu>0$) does not hold for the evolution of birds. However, there is an important qualitative difference between these two grooves: as a function of $\beta$, the birds groove grows considerably slower than the mammals groove. This indicates that for a given diffusion bias $\mu$, a significantly larger extinction parameter $\beta$ is required to produce a comparably realistic mass distribution. In other words, for a particular body mass $M$, the risk of extinction for an avian species should be substantially higher than for a terrestrial mammal species of the same size.

Finally, taking arbitrarily the case of no within-lineage bias \mbox{($\mu=0$)} and using the fitted value of $\beta$, we see that the predicted mass distribution for birds is generally in good agreement with the empirical data (Fig.~\ref{fig:birds}a). The deviations in the left and right tails may make interesting objects for future study. For instance, the deviations in the left tail persist across the groove of minimum wKS values, even for unrealistically large values of $\mu$. This behavior lends some support to the hypothesis that the left tail's particular shape is caused by strong physiological constraints on body mass evolution near the avian lower limit~\citep{stanley:1973}, which are not included in our model. Additionally, the predicted right tail may fit better, with a slightly smaller estimate of $\beta$, were recently extinct species included, such as the Giant Moa ({\em Dinornis robustus}, $250\kg$).

\section*{Discussion and Conclusions}

With respect to birds, our analysis leads us to make several concrete predictions about their evolution. \mbox{(i) As} with terrestrial mammals, avian evolution near the lower limit $\mmin$ is highly constrained, but in ways that that differ to some degree from those faced by mammals near their lower limit. This implies that avian species evolving near this limit should, more often than not, leave larger descendants $\langle\ln\lambda\rangle>0$ and that the strength of this tendency should increase strongly as $M\to\mmin$. \mbox{(ii) The} body masses of large avian species ($M>20\g$), like those for terrestrial mammals, are constrained mainly by extinction risks that increase progressively with body mass. And, \mbox{(iii) avian} species with a mass $M$ face a significantly greater long-term risk of extinction than do mammalian species of the same size. This difference in extinction risk is large enough that an avian species with a mass of roughly $10^{5}\g$ should experience an extinction risk comparable to that of a mammalian species roughly 100 times larger ($10^{7}\g$).

From a more conceptual perspective, the similarity of the results for extant birds and Recent terrestrial mammals suggests that their evolutionary histories, in terms of the processes that govern the variation of species body sizes over evolutionary timescales, are fundamentally the same. The similarity of these distributions to those of other taxonomic groups suggests that this explanation may be universal, although further empirical work is necessary to substantiate this hypothesis. Indeed, recent analyses of fossil data for dinosaurs~\citep{carrano:2006} seems to support this view.

On the other hand, although the agreement of the simple diffusion-reaction model given in Eq.~\eqref{eq:simple}, when appropriately parameterized, and the observed mass distributions of Recent terrestrial mammals and extant birds is quite good, the model is obviously incomplete in many ways. In particular, the number of small-bodied species for both mammals and birds is overestimated by the model, which is likely because we omitted the increased positive bias toward larger sizes identified by Clauset and Erwin. Additionally, speciation and extinction events are assumed to be independent, and thus the model cannot explain fluctuations in the mass distribution caused by phylogenetically correlated extinctions, where entire genera or niches are wiped out by ecological or environmental changes.

We conclude by noting that the model's good agreement with data suggests that it may be a useful way to establish null-expectations in the study of general trends in body mass evolution (much like diffusion models in population genetics~\citep{hartl:clark:1989}) in the absence of factors such as inter-specific competition, population dynamics, geography, predation, etc. For instance, the fully time-dependent formulation in Eq.~\eqref{eq:cxt} could be used to correctly determine the significance of statistical trends in body masses over evolutionary time~\citep{alroy:2000b}. Additionally, given the strong correlations between body mass and other species characteristics, this model of body mass evolution may provide a way to unify certain aspects of ecology and evolution.

\newpage
\acknowledgments
The authors thank D. H. Erwin, D. Krakauer and J. Wilkins for helpful conversations, and J. Alroy, A. Boyer and F. Smith for kindly sharing data. SR gratefully acknowledges support from NSF grant DMR0535503, and DJS support from NSF grant DMR0404507 and the GRM Fellowship at UCLA. This work was supported in part by the Santa Fe Institute.

\end{document}